\documentclass[a4paper]{mn2e}
\usepackage{natbib_jrm,times}
\usepackage{graphicx}
\bibpunct[, ]{(}{)}{;}{a}{}{,}

\def \aj {AJ}
\def \mnras {MNRAS}
\def \pasp {PASP}
\def \apj {ApJ}

\def \apjl {ApJL}
\def \aap {A\&A}
\def \nat {Nature}
\def \araa {ARAA}

\def \apss {Ap\&SS}

\def\lesssim{\mathrel{\hbox{\rlap{\hbox{\lower4pt\hbox{$\sim$}}}\hbox{$<$}}}}
\def\gtrsim{\mathrel{\hbox{\rlap{\hbox{\lower4pt\hbox{$\sim$}}}\hbox{$>$}}}}

\long\def\symbolfootnote[#1]#2{\begingroup%
\def\thefootnote{\fnsymbol{footnote}}\footnote[#1]{#2}\endgroup} 

\begin{document}
\title[Spectropolarimetry of SN~2012fr]{Spectropolarimetry of the Type Ia Supernova 2012fr\thanks{Based on observations made with ESO Telescopes at the Paranal Observatory, under program 290.D-5009 and 290.D-5006.}}
\author[Maund et al.]{
\parbox[t]{\textwidth}{\raggedright
J.R. ~Maund$^{1,2,3}$\thanks{Email: j.maund@qub.ac.uk}, J. Spyromilio$^{4}$, P.A. H\"{o}flich$^{5}$, J.C. Wheeler$^{6}$, D. Baade$^{4}$,\\ A. Clocchiatti$^{7}$, F. Patat$^{4}$,  E. Reilly$^{1}$, L. Wang$^{8}$ \& P. Zelaya$^{7}$}
\vspace*{6pt}\\
$^{1}$ Astrophysics Research Centre, School of Mathematics and Physics, Queen's University Belfast, Belfast, BT7 1NN, Northern Ireland, U.K.\\
$^{2}$ Dark Cosmology Centre, Niels Bohr Institute, University of Copenhagen, Juliane Maries Vej 30, 2100 Copenhagen, DK.\\
$^{3}$ Royal Society Research Fellow\\
$^{4}$ ESO - European Organisation for Astronomical Research in the Southern Hemisphere, Karl-Schwarzschild-Str.\ 2, 85748 Garching b.\ M\"unchen, Germany\\
$^{5}$ Department of Physics, Florida State University, Tallahassee, Florida 32306-4350, U.S.A.\\ 
$^{6}$ Department of Astronomy and McDonald Observatory, The University of Texas, 1 University Station C1402, Austin, Texas 78712-0259, U.S.A.\\ 
$^{7}$ Departamento de Astronomia y Astrofisica, Pontificia Universidad Catolica Casilla 306, Santiago 22, Chile\\ 
$^{8}$ Department of Physics, Texas A\&M University, College Station, Texas 77843-4242, U.S.A.}
\maketitle
\begin{abstract}
Spectropolarimetry provides the means to probe the 3D geometries of
Supernovae at early times.  We report spectropolarimetric observations
of the Type Ia Supernova 2012fr at four epochs: -11, -5, +2 and +24
days, with respect to $B$-lightcurve maximum.  SN~2012fr is a normal
Type Ia SN, similar to SNe 1990N, 2000cx and 2005hj (that all exhibit
low velocity decline rates for the principal Si {\sc ii} line).  The
SN displays high velocity components at -11 days that are highly
polarized.  The polarization of these features decreases as they become
weaker from -5 days.  At +2 days, the polarization angles of the low
velocity components of silicon and calcium are identical and oriented
at $90^{\circ}$ relative to the high velocity Ca component.  In
addition to having very different velocities, the high and low
velocity Ca components have orthogonal distributions in the plane of
the sky.  The continuum polarization for the SN at all four epochs is
low $<0.1\%$.  We conclude that the low level of continuum
polarization is inconsistent with the merger-induced explosion
scenario.  The simple axial symmetry evident from the polarization
angles of the high velocity and low velocity Ca components, along with
the presence of high velocity components of Si and Ca, is perhaps more consistent
with the pulsating delayed detonation model.  We predict that, during
the nebular phase, SN~2012fr will display blue-shifted emission lines
of Fe-group elements.
\end{abstract}
\begin{keywords} supernovae:general -- supernovae:individual:2012fr
\end{keywords}

\section{Introduction}
\label{intro}
Type Ia Supernovae (SNe) are energetic events resulting from the
thermonuclear explosions of carbon-oxygen Chandrasekhar mass white
dwarfs (WDs), in either single or double degenerate progenitor systems
\citep{branIa}.  The exact nature of the explosion mechanism behind
Type Ia SNe is unknown, but a number of theoretical models have been
proposed: deflagrations
\citep{2004PhRvL..92u1102G,2006A&A...453..203R}, detonations
\citep{1969Ap&SS...5..180A}, delayed detonations
\citep{1991A&A...245..114K}, pulsating delayed detonations
\citep{1991A&A...245L..25K,1993A&A...270..223K} or the mergers of two WDs
\citep{1984ApJ...277..355W,2012ApJ...747L..10P}.  A key
differentiating factor between these different models, with different
physical considerations, is the geometry of the resulting
explosion. Spectropolarimetry permits the direct observation of the 3D
geometries of SN explosions and provides, therefore, a direct probe of
the underlying physics behind these explosions \citep[for a review,
  see][]{2008ARA&A..46..433W}.

In general, Type Ia SNe are characterized by low levels of
polarization, that decrease as time and the depth into the ejecta
increase.  Previous studies have shown that photometric and
spectroscopic properties of Type Ia SNe are intimately tied to the
geometry of these events, through the correlation of the polarization
(specifically of the Si {\sc ii} $\lambda 6355$ feature $p_{Si II}$) with the
lightcurve decline parameter $\Delta m_{15}(B)$
\citep{2007Sci...315..212W} and the decline rate of the velocity at
the absorption minimum $\dot{v}_{Si II}$ \citep{2010ApJ...725L.167M}.

Here we report early spectropolarimetric observations of the Type Ia
SN 2012fr.  SN~2012fr was discovered by \citet{2012CBET.3277....1K} on
2012 Oct 27.05 (UT) in the galaxy NGC~1365 ($3\arcsec$ East and
$52\arcsec$ North of the nucleus).  Based on previous observations of
NGC~1365, the SN was discovered $<3$d post-explosion
\citep{2012CBET.3277....1K}.  \citet{2012CBET.3277....2C} and
\citet{2012CBET.3277....3B} spectroscopically classified SN~2012fr as
being a normal Type Ia SN.  The heliocentric recessional velocity for
NGC~1365\footnote{Quoted from the NASA/IPAC Extragalactic Database -
  http://ned.ipac.caltech.edu} is $1636\,\mathrm{km\,s^{-1}}$.
Contreras et al. (2013, in prep.) report that the B-band light curve
maximum occurred at 2012 Nov 12.04 and measured $\Delta
m_{15}(B)=0.80$.
\section{Observations and Data Reduction}
\label{sec:obs}
Spectropolarimetric observations of SN~2012fr were conducted using the
European Southern Observatory (ESO) Very Large Telescope (VLT) Antu
Unit Telescope and the Focal Reducer and low dispersion Spectrograph
(FORS; \citealt{1998Msngr..94....1A}).  The observations were made at
four separate epochs, and a log of the observations is presented in
Table \ref{tab:obs}.  The data were acquired using the $300V$ grism
and $GG435$ order separation filter (to prevent second order
contamination in the red portion of the spectrum), providing a
wavelength range of $4500-9300{\mathrm \AA}$ with resolution $\sim
11.8{\mathrm \AA}$ at $6000{\mathrm \AA}$.  At the +2d, observations
were conducted with both the half-wavelength and quarter wavelength
retarder plates, to measure the linear ($Q$ and $U$) and circular
($V$) Stokes parameters, respectively.  The data were reduced using
IRAF\footnote{IRAF is distributed by the National Optical Astronomy
  Observatory, which is operated by the Association of Universities
  for Research in Astronomy (AURA) under cooperative agreement with
  the National Science Foundation.} following the method presented by
\citet{maund05bf}.  The reduction procedure was identical for
observations conducted with half- and quarter-wavelength retarder
plates.  Flux spectra of the target SN were calibrated using
observations of a flux standard acquired with the polarimetry optics
in place.  The Stokes parameters were rebinned to $15{\mathrm  \AA}$,
slightly larger than one resolution element, to increase the level of
signal to noise.

   \begin{table}
      \caption[]{\label{tab:obs}Log of Spectropolarimetric Observations}
  \begin{tabular}{ccccc}
\hline\hline
Object                      &    Date       &   Exposure    &   Mean      & Epoch$^{\star}$    \\
                            &    (UT)       &     (s)       &  Airmass    &  (days)        \\
             
\hline
SN~2012fr                   & 2012 Nov 1.07  & $4\times 170$ &   1.56    & -11\\
\\
LTT1020$^{\ddagger}$          & 2012 Nov 7.06  & $30$          &   1.20 \\
SN~2012fr                   & 2012 Nov 7.24  & $4\times 100$ &   1.03   & -5 \\
\\
LTT1020$^{\dagger}$$^{\ddagger}$& 2012 Nov 14.19 & $4\times 30$  & 1.09 \\   
LTT1020$^{\ddagger}$          & 2012 Nov 14.19 & $8\times 30$  &   1.08        \\
SN~2012fr                   & 2012 Nov 14.22 & $4\times 100$ &   1.03   &  +2\\
SN~2012fr                   & 2012 Nov 14.23 & $4\times 80$  &   1.04   &  +2\\
SN~2012fr$^{\dagger}$         & 2012 Nov 14.23 & $4\times 30$  &   1.04   & +2 \\
\\
SN~2012fr                   & 2012 Dec 06.31 & $16\times 125$&2.02 \\
G94-48$^{\ddagger}$           & 2012 Dec 07.01 & $4 \times 60$ & 1.49 &+24 \\
\hline\hline

\end{tabular}\\
$^{\star}$ Relative to the maximum of the $B$-band lightcurve on 2012 Nov 12.04\\
$^{\dagger}$ Observations conducted with the quarter wavelength retarder plate.\\
$^{\ddagger}$ Flux standard.\\
\end{table}
\section{Results}
\label{sec:res} 
\begin{figure}
\includegraphics[width=8.5cm]{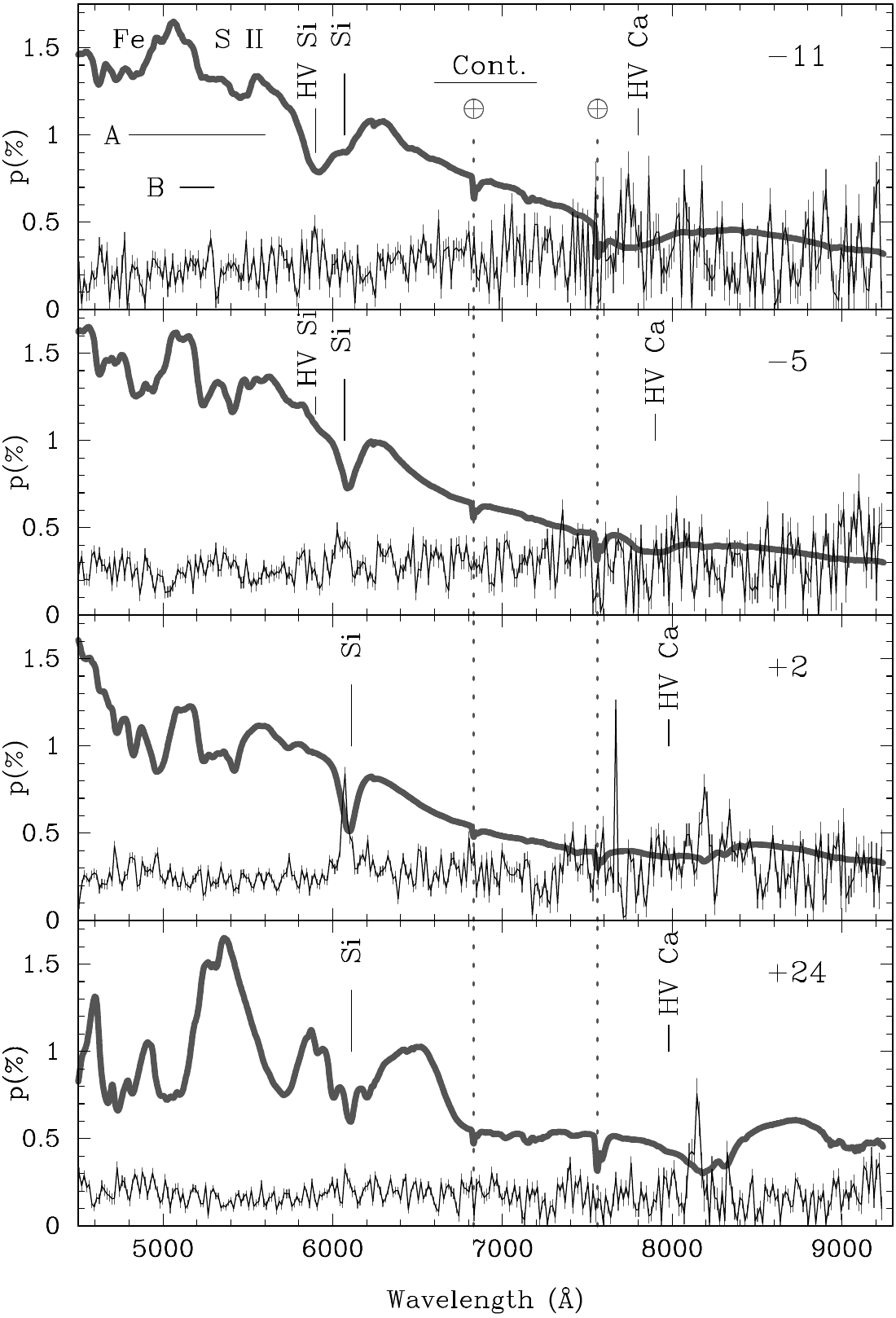}
\caption{Polarization spectra of SN~2012fr at -11, -5, +2 and +24d
  relative to $B$-lightcurve maximum.  A scaled flux spectrum
  ($\mathrm{erg \, s^{-1} \, cm^{-2} \, s^{-1}}$) is shown as the
  thick line.  All spectra have been corrected for the recessional
  velocity of the host galaxy.  The polarization spectra have not been
  corrected for the interstellar polarization component.  The
  wavelength ranges used for the two estimates of the interstellar
  polarization (labelled ``A'' and ``B'') and for measuring the
  continuum polarization (labelled ``Cont.'') are indicated by the
  horizontal bars.  The principal telluric lines are indicated by the
  $\oplus$ symbol.}
\label{fig:res:pol}
\end{figure}

The polarization and flux spectra for SN~2012fr at the four
observational epochs are presented in Figure \ref{fig:res:pol}. We
compared the flux spectra at each epoch with those of other Type Ia
SNe using {\it
  GELATO}\footnote{http://gelato.tng.iac.es/}\citep{2008A&A...488..383H}.The
first two flux spectra appear similar to SN~1990N at -13 and -7d
(relative to $B$-band maximum; \citealt{1991ApJ...371L..23L}) and to
the ``Branch normal'' SN 2005hj \citep{2007ApJ...666.1083Q}; whilst
the later spectra appear similar to 2000cx
\citep{2001PASP..113.1178L}.

At the first epoch, the Si {\sc ii} $\lambda 6355$ line is
composed of two separate absorption components, a high velocity (HV)
component at $-20\,000\,{\mathrm{km\,s^{-1}}}$ and a low velocity (LV)
component at $-12\,800\,\mathrm{km\,s^{-1}}$.  In the later epochs, Si
{\sc ii} is dominated by the narrow LV component, although HV
absorption is discernible at the second epoch.  Based on the measured
decline of the velocity of the Si {\sc ii} LV component over the four epochs ($\dot v_{Si II}= 25\pm 7\,
\mathrm{km\,s^{-1}\,d^{-1}}$), SN~2012fr is a member of the Low
Velocity Gradient (LVG) group of Type Ia SNe \citep{2005ApJ...623.1011B}.
The HV component of Si {\sc ii} appears detached from
the LV component, and we find  the velocity at the blue edge of
the absorption trough at -11d is the same as the velocity measured at
absorption minimum for the Ca {\sc ii} Infrared Triplet (IR3), as observed by \citet{2007ApJ...666.1083Q} in spectra of SN~2005hj.

At the first two epochs, the Ca {\sc ii} IR3 profile (shown in Figure
\ref{fig:res:ca2}) displays a significant HV component ($-28\,600$ and
$-23\,500\,\mathrm{km\,s^{-1}}$, for the first and second epochs
respectively), with three weaker LV components (labeled A, B and C in
Fig. \ref{fig:res:ca2}).  The HV component appears in the last two
epochs, but as a much weaker absorption, while the strongest
absorption arises from the ``A'' LV component.  At +2d, we measure
velocities of the three LV components to be $-10\,900$, $-8\,200$ and
$-8\,000\,\mathrm{km\,s^{-1}}$, relative to Ca {\sc ii
}$\lambda\lambda 8498,8542,8662$, respectively.  The ``saw tooth''
pattern of absorption components (+2d) is similar to that observed for
SN 2000cx at +7d post-maximum \citep{2001PASP..113.1178L}.  We suggest
that the LV components arise from the partially resolved individual
lines of which the IR3 is composed.

At each epoch the polarization spectrum is characterized by a
relatively low level of polarization $0.3-0.4\%$.  Although peaks in
the polarization spectra may be associated with the blue Fe lines and
the S {\sc ii} absorption features ($\mathrm{5200-5400\AA}$), the most
significantly polarized features are the Si {\sc ii} and Ca {\sc ii}
IR3 lines.  At the first two epochs, the apparent rise in polarization
at longer wavelengths coincides with the increasing uncertainty on the
polarization commensurate with the decreasing signal-to-noise.  The
apparent polarization at $\mathrm{\sim 7600\AA}$ is due to the
significant telluric absorption around that wavelength and is not
intrinsic to the SN.  On the Stokes $Q-U$ plane (see Figure
\ref{fig:res:squ}), the data at all epochs are characterized by a
central concentration of points, with significant departures at
wavelengths associated with Si {\sc ii} and Ca {\sc ii} features.

Stokes $V$ was measured at $+2\mathrm{d}$ to be $0\pm0.2\%$,
consistent with there being no circular polarization over the entire
wavelength range; as was the case with the previous reported
broad-band polarimetric measurement of the Type I SN 1972E
\citep{1972Natur.238..452W} and spectropolarimetry of Type Ia SN 1983J
\citep{1984MNRAS.210..839M}.  The equipment used for the observations made of SN 1972E, one month after maximum light, is not identified.  SN 1983J was observed around maximum light with a 4-m telescope and no solid-state detector.  Our VLT CCD data for SN 2012fr, at maximum light, should set the tightest limits to date on circular polarization of any Type Ia SN (or even yield a marginal detection).  We will pursue this in a separate study.

\section{Analysis}
\label{sec:ana}
The correct interpretation of the intrinsic polarization of SNe
requires the removal of the interstellar polarization (ISP).  We
estimated the degree of the ISP directly from the observations, under
the assumption that regions of the SN spectra are intrinsically
unpolarized due to line blanketing of Fe lines
\citep{2001ApJ...556..302H,2006PASP..118..722C,my2005hk,2009A&A...508..229P}.
We identify two regions in the spectrum of SN~2012fr that are likely to
be intrinsically depolarized due to multiple overlapping Fe lines at
the first three epochs: regions A (${\mathrm 4800-5600\AA}$) and B
($\mathrm{5100-5300\AA}$), as shown in Fig. \ref{fig:res:pol}. At
+24d it is apparent that lines in these regions have ceased to be
blended and we see strong individual line profiles, so it was not included in the determination of the ISP.  The weighted
average Stokes parameters over these wavelength ranges were used as a
measure of the ISP:
$<Q_{ISP}^{A}>=0.24\pm0.04\%$,$<U_{ISP}^{A}>=0.00\pm0.04\%$ and
$<Q_{ISP}^{B}>=0.24\pm0.04\%$,$<U_{ISP}^{B}>=0.00\pm0.04\%$.  These correspond to $p_{ISP}=0.24\pm0.01\%$ and $\chi_{ISP}=0.0\pm4.9^{\circ}$, roughly aligned with the position
angle of SN~2012fr with respect to the nucleus of NGC~1365.

Under the assumption that the wavelength region $\mathrm{6600-7200
  \AA}$ is representative of the continuum
\citep{2009A&A...508..229P}, we the weighted average continuum
polarization (after correction for the ISP) to be: $0.06\pm0.12\%$,
$0.05\pm0.09\%$, $0.03\pm0.09\%$ and $0.06\pm0.07\%$ at -11, -5, +2
and +24d respectively, consistent with null polarization.  The low
level of continuum polarization is consistent with a spherical
photosphere (with axial ratio $>0.95$;
\citealt{1991A&A...246..481H}). After correction for the ISP, the only
lines remaining with significant polarization are the Si {\sc ii}
$\lambda 6355$ and Ca {\sc ii} IR3 lines.

At -11d, the polarization signal associated with Si {\sc ii} is
dominated by the HV component at $0.40\pm0.06\%$, but as the optical
depth of this component decreases we no longer observe it to be
polarized by -5d; whilst the polarization signal of the LV component
increases.  On the Stokes $Q-U$ plane (see Figure \ref{fig:res:squ}),
the polarization signals of the HV component (observed at -11d) and
the LV component (at -5d) are found to be separated by $\Delta \chi
\approx 20 ^{\circ}$ (which is evidence that the HV and LV components
are physically separated, both in radial velocity and on the sky).
Although the degree of polarization of the  LV component increases from
$0.30\pm0.05\%$ (at -5d) to $0.65\pm0.04\%$ (at +2d), the polarization
angle remains the same, implying that the LV Si {\sc ii} line forming
region is a single continuous region covering the photosphere.  At
+24d, we measure a reduced level of polarization of $0.20\pm0.03\%$
associated with the now weaker Si {\sc ii} absorption, accompanied by
a rotation of the polarization angle through $\approx 70^{\circ}$.

\begin{figure*}
\includegraphics[width=16.5cm]{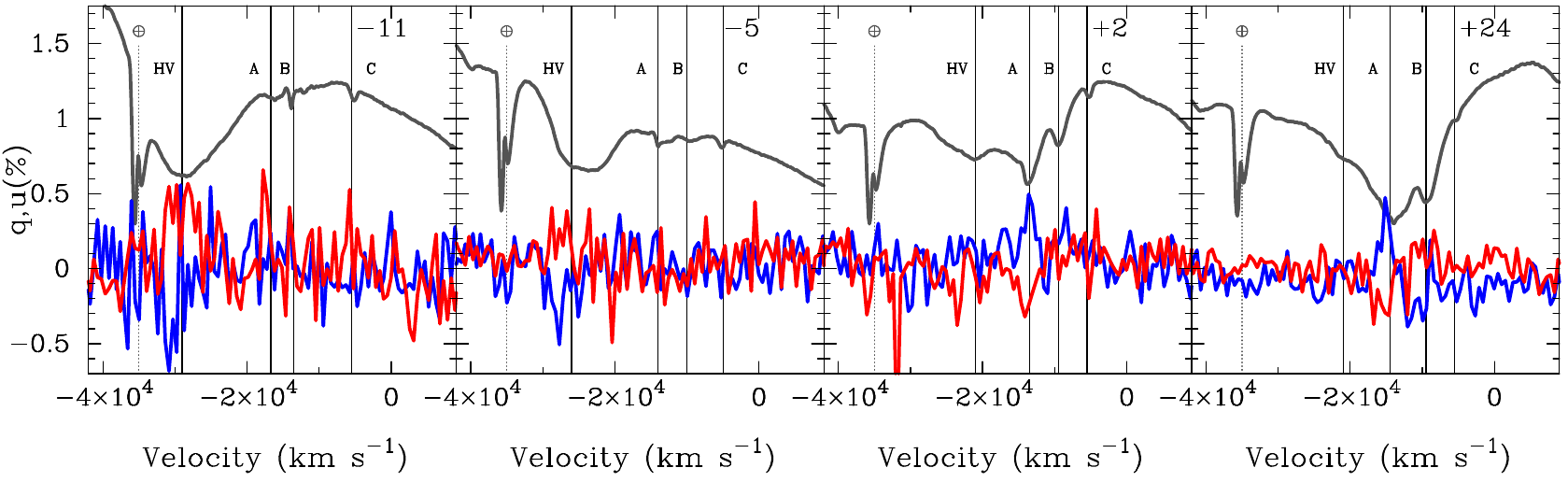}
\caption{The Stokes $Q$ (blue) and $U$ (red) parameters across the Ca
  {\sc ii} IR3 at -11, -5, +2 and +24d, relative to $B$-lightcurve
  maximum.  The Stokes parameters have been corrected for the ISP.
  The scaled flux spectrum is shown by the thick line.  The velocity
  scale is defined relative to average wavelength of the Ca {\sc ii}
  IR3 (${\mathrm 8579\AA}$).  The high velocity component is labeled ``HV'',
  while the three LV components are labeled A-C.  The principal
  telluric line is indicated by the $\oplus$ symbol.}
\label{fig:res:ca2}
\end{figure*}

\begin{figure*}

\includegraphics[width=16.5cm]{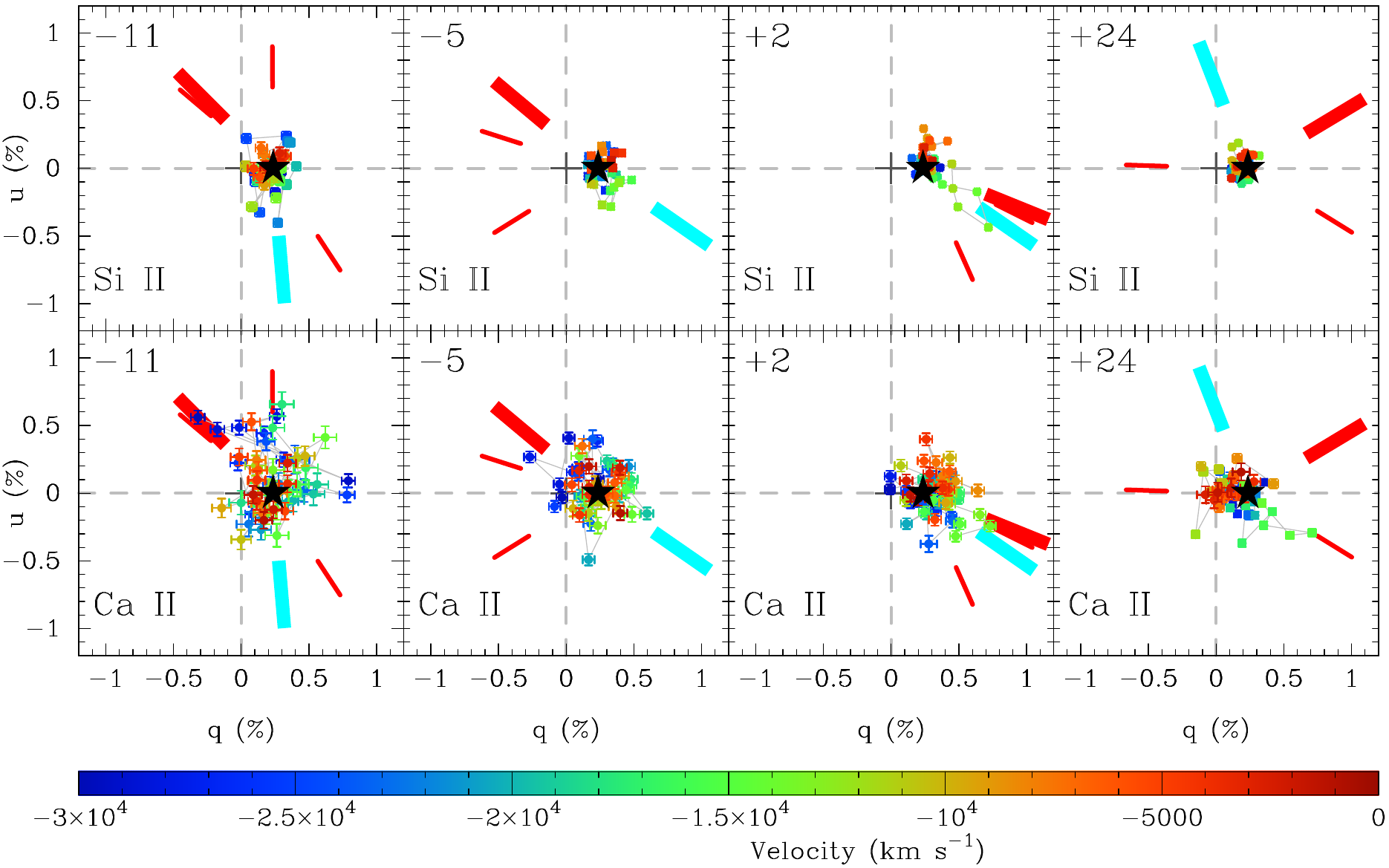}
\caption{The polarization for the Si {\sc ii} $\lambda 6355$ (top row)
  and the Ca {\sc ii} IR3 (bottom row) lines on the Stokes $Q-U$ plane
  at -11, -5, +2 and +24d, relative to $B$-lightcurve maximum.  The
  points are colour coded according to the velocity relative to the
  respective rest wavelengths for the two features.  The data are not
  corrected for the ISP (the location of which is indicated by the
  $\star$).  The thick radial lines indicate the polarization angles
  (on the Stokes $Q-U$ plane) for Si {\sc ii} (light blue) and the HV
  component of Ca {\sc ii} IR3 (red) (the LV components are indicated
  by narrow red radial lines).}
\label{fig:res:squ}
\end{figure*}

At the first epoch, the HV component of Ca {\sc ii} shows the largest
polarization of $0.85\pm0.13\%$; by -5d the degree of polarization
decreases to $0.31\pm0.08\%$ (see Figure \ref{fig:res:ca2}).  At +2d,
the polarization signal associated with Ca {\sc ii} is dominated by
the LV component with $0.54\pm0.07\%$ (specifically the absorption ``A''
in Fig. \ref{fig:res:ca2}).  From Fig. \ref{fig:res:ca2} it is clear
that the Stokes parameters associated with the HV component at earlier
epochs are inverted with respect to the Stokes parameters of the LV
components observed late (i.e. the signs of Stokes $Q$ and $U$
parameters for the HV component are opposite to those of the LV
component); this is also visible as a rotation of the polarization
angle of the Ca {\sc ii} features on Fig. \ref{fig:res:squ}.  This
implies that the HV and LV components arise from line forming regions
that are not only kinematically distinct, but also orthogonal to each
other on the plane of the sky.  At +2 and +24d, the polarization
angles of the strong Ca {\sc ii} absorption (arising from the LV
component) are aligned with the highly polarized Si {\sc ii} feature
observed at +2d, suggesting a common line forming region; however, the
velocities of the LV Ca {\sc ii} components are lower and decline
faster than for Si {\sc ii} $\lambda 6355$.  At +24d, the polarization
signal at the wavelength of the HV component is dominated by the
polarization of the underlying LV component.

\section{Discussion \& Conclusions}
\label{sec:disc}
We have presented early spectropolarimetric observations of the Type
Ia SN~2012fr; finding the degree of polarization to be generally low,
except for significant line polarization associated with Si {\sc ii}
and Ca {\sc ii}.  At early times we see evidence for kinematically
distinct HV components, which weaken significantly by +2d.
Simultaneously, the strength of the LV components increases and their
respective polarizations also increase, but at different polarization
angles than the HV components.  The low decline rate of the LV Si {\sc
  ii} velocity and the degree of polarization measured for the line at
-5d adheres closely to the relationship established by
\citet{2010ApJ...725L.167M} (see their Figure 1) between these two
parameters; and closely matches the values measured for the 1991T-like
SN 1997br \citep{1999AJ....117.2709L,2007Sci...315..212W}.  The degree
of Si {\sc ii} polarization and the measured $\Delta m_{15}(B)$ also
follows the correlation found by \citet{2007Sci...315..212W}.  In
comparison with the \citeauthor{2010ApJ...725L.167M} sample, SN~2012fr
is the only SN with $\Delta m_{15}(B) < 1$ that also exhibits low
$p_{Si II}$ and belongs to the LVG group.  Based on the previous
correlation established between Si {\sc ii} polarization and the
characteristics of the late-time nebular spectrum
\citep{2010ApJ...725L.167M,maedapap}, we predict that the profiles of
lines of Fe-group elements in the nebular spectrum of SN~2012fr will
exhibit a blue shift.

The presence of short lived HV Ca {\sc ii} IR3 components in the
spectra of Type Ia SNe is not unusual. \citet{2004ApJ...607..391G}
modeled this feature, in SN~2003du, as arising from a low mass ($\sim
0.02M_{\odot}$) shell of primordial material from the circumstellar medium.  Both
HV Si {\sc ii} and Ca {\sc ii} were observed in the spectra of
SN~2005hj \citep{2007ApJ...666.1083Q}.  The presence of these two
features in a similar velocity space demands the presence of a large
mass of SN processed material at high velocities, leading
\citeauthor{2007ApJ...666.1083Q} to suggest an explosive origin for
the shell through pulsating delayed detonations (PDDs;
\citealt{1991A&A...245..114K,1993A&A...270..223K}) or
mergers. The 3D signatures of these mechanisms, however, will be very
different.  Approximate spherical symmetry is expected to be maintained in
the PDD model (although this must be verified with multidimensional
simulations), but large scale asymmetries are expected in mergers.  The overall low level of continuum polarization observed for
SN~2012fr, however, is inconsistent with the asymmetric nickel
distributions predicted by \citet{2012ApJ...747L..10P} for violent mergers.  The apparent
orthogonality of the line forming regions for the HV and LV Ca {\sc
  ii} components observed for SN~2012fr, on the plane of the sky, is
evidence for a simple axial symmetry.  We also note that the slow rise
time of the light curve of SN~2012fr and the low decline rate of the
Si {\sc ii} velocity are consistent with the predictions of the PDD
model \citep{1993A&A...270..223K,1996ApJ...457..500H}.

The spectroscopic and photometric similarities between the low $\dot
v_{Si_{II}}$, low $\Delta m_{15}(B)$ SNe, such as 1990N, 2000cx and
2005hj, with SN 2012fr, suggest that these events may be similar events observed at similar orientations (possibly consistent with the PDD model).

\section*{Acknowledgments} 
We thank the ESO Director General for awarding discretionary time for
the observations and the observers at Paranal for conducting such
exquisite observations.  We thank Carlos Contreras and Jeffery
Silverman for early notification of the results of their photometric
monitoring of SN~2012fr.  The research of JRM is supported through a
Royal Society University Research Fellowship.  The work of JCW is
supported in part by NSF Grant AST-1109801.  AC \& PZ acknowledge
support from Iniciativa Cientifica Milenio (MINECON, Chile) through
the Millennium Center for Supernova Science (P10-064-F) and Basal
grant CATA PFB 06/09 from CONICYT, Chile.

\bibliographystyle{mn2e}

\begin{thebibliography}{}

\bibitem[\protect\citeauthoryear{{Appenzeller}, {Fricke}, {Furtig}, {Gassler},
  {Hafner}, {Harkl}, {Hess}, {Hummel} \& {et~al.}}{{Appenzeller}
  et~al.}{1998}]{1998Msngr..94....1A}
{Appenzeller} I.,  {Fricke} K.,  {Furtig} W.,  {Gassler} W.,  {Hafner} R.,
  {Harkl} R.,  {Hess} H.-J.,  {Hummel} W.,    {et~al.} 1998, The Messenger, 94,
  1

\bibitem[\protect\citeauthoryear{{Arnett}}{{Arnett}}{1969}]{1969Ap&SS...5..180A}
{Arnett} W.~D.,  1969, \apss, 5, 180

\bibitem[\protect\citeauthoryear{{Benetti}, {Cappellaro}, {Mazzali}, {Turatto},
  {Altavilla}, {Bufano}, {Elias-Rosa}, {Kotak}, {Pignata}, {Salvo} \&
  {Stanishev}}{{Benetti} et~al.}{2005}]{2005ApJ...623.1011B}
{Benetti} S.,  {Cappellaro} E.,  {Mazzali} P.~A.,  {Turatto} M.,  {Altavilla}
  G.,  {Bufano} F.,  {Elias-Rosa} N.,  {Kotak} R.,  {Pignata} G.,  {Salvo} M.,
    {Stanishev} V.,  2005, \apj, 623, 1011

\bibitem[\protect\citeauthoryear{{Branch}, {Livio}, {Yungelson}, {Boffi} \&
  {Baron}}{{Branch} et~al.}{1995}]{branIa}
{Branch} D.,  {Livio} M.,  {Yungelson} L.~R.,  {Boffi} F.~R.,    {Baron} E.,
  1995, \pasp, 107, 1019

\bibitem[\protect\citeauthoryear{{Buil}}{{Buil}}{2012}]{2012CBET.3277....3B}
{Buil} C.,  2012, Central Bureau Electronic Telegrams, 3277, 3

\bibitem[\protect\citeauthoryear{{Childress}, {Zhou}, {Tucker}, {Bayliss},
  {Scalzo}, {Yuan} \& {Schmidt}}{{Childress}
  et~al.}{2012}]{2012CBET.3277....2C}
{Childress} M.,  {Zhou} G.,  {Tucker} B.,  {Bayliss} D.,  {Scalzo} R.,  {Yuan}
  F.,    {Schmidt} B.,  2012, Central Bureau Electronic Telegrams, 3277, 2

\bibitem[\protect\citeauthoryear{{Chornock}, {Filippenko}, {Branch}, {Foley},
  {Jha} \& {Li}}{{Chornock} et~al.}{2006}]{2006PASP..118..722C}
{Chornock} R.,  {Filippenko} A.~V.,  {Branch} D.,  {Foley} R.~J.,  {Jha} S.,
  {Li} W.,  2006, \pasp, 118, 722

\bibitem[\protect\citeauthoryear{{Gamezo}, {Khokhlov} \& {Oran}}{{Gamezo}
  et~al.}{2004}]{2004PhRvL..92u1102G}
{Gamezo} V.~N.,  {Khokhlov} A.~M.,    {Oran} E.~S.,  2004, Physical Review
  Letters, 92, 211102

\bibitem[\protect\citeauthoryear{{Gerardy}, {H{\"o}flich}, {Fesen}, {Marion},
  {Nomoto}, {Quimby}, {Schaefer}, {Wang} \& {Wheeler}}{{Gerardy}
  et~al.}{2004}]{2004ApJ...607..391G}
{Gerardy} C.~L.,  {H{\"o}flich} P.,  {Fesen} R.~A.,  {Marion} G.~H.,  {Nomoto}
  K.,  {Quimby} R.,  {Schaefer} B.~E.,  {Wang} L.,    {Wheeler} J.~C.,  2004,
  \apj, 607, 391

\bibitem[\protect\citeauthoryear{{Harutyunyan}, {Pfahler}, {Pastorello},
  {Taubenberger}, {Turatto}, {Cappellaro}, {Benetti}, {Elias-Rosa},
  {Navasardyan}, {Valenti}, {Stanishev}, {Patat}, {Riello}, {Pignata} \&
  {Hillebrandt}}{{Harutyunyan} et~al.}{2008}]{2008A&A...488..383H}
{Harutyunyan} A.~H.,  {Pfahler} P.,  {Pastorello} A.,  {Taubenberger} S.,
  {Turatto} M.,  {Cappellaro} E.,  {Benetti} S.,  {Elias-Rosa} N.,
  {Navasardyan} H.,  {Valenti} S.,  {Stanishev} V.,  {Patat} F.,  {Riello} M.,
  {Pignata} G.,    {Hillebrandt} W.,  2008, \aap, 488, 383

\bibitem[\protect\citeauthoryear{{Hoeflich} \& {Khokhlov}}{{Hoeflich} \&
  {Khokhlov}}{1996}]{1996ApJ...457..500H}
{Hoeflich} P.,  {Khokhlov} A.,  1996, \apj, 457, 500

\bibitem[\protect\citeauthoryear{{H\"{o}flich}}{{H\"{o}flich}}{1991}]{1991A&A...246..481H}
{H\"{o}flich} P.,  1991, \aap, 246, 481

\bibitem[\protect\citeauthoryear{{Howell}, {H{\"o}flich}, {Wang} \&
  {Wheeler}}{{Howell} et~al.}{2001}]{2001ApJ...556..302H}
{Howell} D.~A.,  {H{\"o}flich} P.,  {Wang} L.,    {Wheeler} J.~C.,  2001, \apj,
  556, 302

\bibitem[\protect\citeauthoryear{{Khokhlov}, {Mueller} \&
  {Hoeflich}}{{Khokhlov} et~al.}{1993}]{1993A&A...270..223K}
{Khokhlov} A.,  {Mueller} E.,    {Hoeflich} P.,  1993, \aap, 270, 223

\bibitem[\protect\citeauthoryear{{Khokhlov}}{{Khokhlov}}{1991a}]{1991A&A...245..114K}
{Khokhlov} A.~M.,  1991a, \aap, 245, 114

\bibitem[\protect\citeauthoryear{{Khokhlov}}{{Khokhlov}}{1991b}]{1991A&A...245L..25K}
{Khokhlov} A.~M.,  1991b, \aap, 245, L25

\bibitem[\protect\citeauthoryear{{Klotz}, {Normand}, {Conseil}, {Parker},
  {Fabrega} \& {Maury}}{{Klotz} et~al.}{2012}]{2012CBET.3277....1K}
{Klotz} A.,  {Normand} J.,  {Conseil} E.,  {Parker} S.,  {Fabrega} J.,
  {Maury} A.,  2012, Central Bureau Electronic Telegrams, 3277, 1

\bibitem[\protect\citeauthoryear{{Leibundgut}, {Kirshner}, {Filippenko},
  {Shields}, {Foltz}, {Phillips} \& {Sonneborn}}{{Leibundgut}
  et~al.}{1991}]{1991ApJ...371L..23L}
{Leibundgut} B.,  {Kirshner} R.~P.,  {Filippenko} A.~V.,  {Shields} J.~C.,
  {Foltz} C.~B.,  {Phillips} M.~M.,    {Sonneborn} G.,  1991, \apjl, 371, L23

\bibitem[\protect\citeauthoryear{{Li}, {Filippenko}, {Gates}, {Chornock},
  {Gal-Yam}, {Ofek}, {Leonard}, {Modjaz}, {Rich}, {Riess} \& {Treffers}}{{Li}
  et~al.}{2001}]{2001PASP..113.1178L}
{Li} W.,  {Filippenko} A.~V.,  {Gates} E.,  {Chornock} R.,  {Gal-Yam} A.,
  {Ofek} E.~O.,  {Leonard} D.~C.,  {Modjaz} M.,  {Rich} R.~M.,  {Riess} A.~G.,
    {Treffers} R.~R.,  2001, \pasp, 113, 1178

\bibitem[\protect\citeauthoryear{{Li}, {Qiu}, {Qiao}, {Zhu}, {Hu}, {Richmond},
  {Filippenko}, {Treffers}, {Peng} \& {Leonard}}{{Li}
  et~al.}{1999}]{1999AJ....117.2709L}
{Li} W.~D.,  {Qiu} Y.~L.,  {Qiao} Q.~Y.,  {Zhu} X.~H.,  {Hu} J.~Y.,  {Richmond}
  M.~W.,  {Filippenko} A.~V.,  {Treffers} R.~R.,  {Peng} C.~Y.,    {Leonard}
  D.~C.,  1999, \aj, 117, 2709

\bibitem[\protect\citeauthoryear{{Maeda}, {Benetti}, {Stritzinger},
  {R{\"o}pke}, {Folatelli}, {Sollerman}, {Taubenberger}, {Nomoto}, {Leloudas},
  {Hamuy}, {Tanaka}, {Mazzali} \& {Elias-Rosa}}{{Maeda}
  et~al.}{2010}]{maedapap}
{Maeda} K.,  {Benetti} S.,  {Stritzinger} M.,  {R{\"o}pke} F.~K.,  {Folatelli}
  G.,  {Sollerman} J.,  {Taubenberger} S.,  {Nomoto} K.,  {Leloudas} G.,
  {Hamuy} M.,  {Tanaka} M.,  {Mazzali} P.~A.,    {Elias-Rosa} N.,  2010, \nat,
  466, 82

\bibitem[\protect\citeauthoryear{{Maund}, {Wheeler}, {Patat}, {Baade}, {Wang}
  \& {H\"{o}flich}}{{Maund} et~al.}{2007}]{maund05bf}
{Maund} J.,  {Wheeler} J.,  {Patat} F.,  {Baade} D.,  {Wang} L.,
  {H\"{o}flich} P.,  2007, \mnras, 381, 201

\bibitem[\protect\citeauthoryear{{Maund}, {H{\"o}flich}, {Patat}, {Wheeler},
  {Zelaya}, {Baade}, {Wang}, {Clocchiatti} \& {Quinn}}{{Maund}
  et~al.}{2010}]{2010ApJ...725L.167M}
{Maund} J.~R.,  {H{\"o}flich} P.,  {Patat} F.,  {Wheeler} J.~C.,  {Zelaya} P.,
  {Baade} D.,  {Wang} L.,  {Clocchiatti} A.,    {Quinn} J.,  2010, \apjl, 725,
  L167

\bibitem[\protect\citeauthoryear{{Maund}, {Wheeler}, {Wang}, {Baade},
  {Clocchiatti}, {Patat}, {H{\"o}flich}, {Quinn} \& {Zelaya}}{{Maund}
  et~al.}{2010}]{my2005hk}
{Maund} J.~R.,  {Wheeler} J.~C.,  {Wang} L.,  {Baade} D.,  {Clocchiatti} A.,
  {Patat} F.,  {H{\"o}flich} P.,  {Quinn} J.,    {Zelaya} P.,  2010, \apj, 722,
  1162

\bibitem[\protect\citeauthoryear{{McCall}, {Reid}, {Bessell} \&
  {Wickramasinghe}}{{McCall} et~al.}{1984}]{1984MNRAS.210..839M}
{McCall} M.~L.,  {Reid} N.,  {Bessell} M.~S.,    {Wickramasinghe} D.,  1984,
  \mnras, 210, 839

\bibitem[\protect\citeauthoryear{{Pakmor}, {Kromer}, {Taubenberger}, {Sim},
  {R{\"o}pke} \& {Hillebrandt}}{{Pakmor} et~al.}{2012}]{2012ApJ...747L..10P}
{Pakmor} R.,  {Kromer} M.,  {Taubenberger} S.,  {Sim} S.~A.,  {R{\"o}pke}
  F.~K.,    {Hillebrandt} W.,  2012, \apjl, 747, L10

\bibitem[\protect\citeauthoryear{{Patat}, {Baade}, {H{\"o}flich}, {Maund},
  {Wang} \& {Wheeler}}{{Patat} et~al.}{2009}]{2009A&A...508..229P}
{Patat} F.,  {Baade} D.,  {H{\"o}flich} P.,  {Maund} J.~R.,  {Wang} L.,
  {Wheeler} J.~C.,  2009, \aap, 508, 229

\bibitem[\protect\citeauthoryear{{Quimby}, {H{\"o}flich} \& {Wheeler}}{{Quimby}
  et~al.}{2007}]{2007ApJ...666.1083Q}
{Quimby} R.,  {H{\"o}flich} P.,    {Wheeler} J.~C.,  2007, \apj, 666, 1083

\bibitem[\protect\citeauthoryear{{R{\"o}pke}, {Gieseler}, {Reinecke},
  {Travaglio} \& {Hillebrandt}}{{R{\"o}pke} et~al.}{2006}]{2006A&A...453..203R}
{R{\"o}pke} F.~K.,  {Gieseler} M.,  {Reinecke} M.,  {Travaglio} C.,
  {Hillebrandt} W.,  2006, \aap, 453, 203

\bibitem[\protect\citeauthoryear{{Wang}, {Baade} \& {Patat}}{{Wang}
  et~al.}{2007}]{2007Sci...315..212W}
{Wang} L.,  {Baade} D.,    {Patat} F.,  2007, Science, 315, 212

\bibitem[\protect\citeauthoryear{{Wang} \& {Wheeler}}{{Wang} \&
  {Wheeler}}{2008}]{2008ARA&A..46..433W}
{Wang} L.,  {Wheeler} J.~C.,  2008, \araa, 46, 433

\bibitem[\protect\citeauthoryear{{Webbink}}{{Webbink}}{1984}]{1984ApJ...277..355W}
{Webbink} R.~F.,  1984, \apj, 277, 355

\bibitem[\protect\citeauthoryear{{Wolstencroft} \& {Kemp}}{{Wolstencroft} \&
  {Kemp}}{1972}]{1972Natur.238..452W}
{Wolstencroft} R.~D.,  {Kemp} J.~C.,  1972, \nat, 238, 452

\end{thebibliography}

\end{document}